\documentclass[preprint,superscriptaddress]{revtex4-1}

\usepackage{graphicx}
\usepackage{amsmath}
\usepackage{amssymb}
\usepackage{setspace}

\begin{document}

\title{Kinky DNA in solution: Small angle
scattering study of a nucleosome positioning sequence.}

\date{\today}

\author{Torben Schindler}
\affiliation{Lehrstuhl f{\"u}r Kristallographie und Strukturphysik,
Friedrich-Alexander-Universit{\"a}t 
Erlangen-N{\"u}rnberg, Staudtstr. 3, D-91058 Erlangen,
Germany}
\author{Adri\'an Gonz\'alez}
\affiliation{Institut Laue-Langevin, 71 avenue des Martyrs, CS 20156, 38042
Grenoble Cedex 9, France}
\affiliation{ICCRAM, University of Burgos, Science and Technology Park, Plaza
Misael Ba{\~{n}}uelos, 09001 Burgos, Spain}
\author{Ramachandran Boopathi}
\affiliation{Universit{\'e} de Lyon,
Laboratoire de Biologie et de Mod{\'e}lisation de la Cellule (LBMC)
CNRS/ENSL/UCBL UMR 5239, Ecole Normale Sup{\'e}rieure de Lyon, 69007 Lyon,
France}
\affiliation{Institut Albert Bonniot, Universit\'{e} de Grenoble Alpes/INSERM
U1209/CNRS UMR 5309, 38042 Grenoble Cedex 9, France}
\author{Marta Marty Roda}
\affiliation{ICCRAM, University of Burgos, Science and Technology Park, Plaza
Misael Ba{\~{n}}uelos, 09001 Burgos, Spain}
\author{Lorena Romero-Santacreu}
\affiliation{ICCRAM, University of Burgos, Science and Technology Park, Plaza
Misael Ba{\~{n}}uelos, 09001 Burgos, Spain}
\author{Andrew Wildes}
\affiliation{Institut Laue-Langevin, 71 avenue des Martyrs, CS 20156, 38042
Grenoble Cedex 9, France}
\author{Lionel Porcar}
\affiliation{Institut Laue-Langevin, 71 avenue des Martyrs, CS 20156, 38042
Grenoble Cedex 9, France}
\author{Anne Martel}
\affiliation{Institut Laue-Langevin, 71 avenue des Martyrs, CS 20156, 38042
Grenoble Cedex 9, France}
\author{Nikos Theodorakopoulos}
\affiliation{Theoretical and Physical Chemistry Institute, National Hellenic
  Research Foundation, Vasileos Constantinou 48, GR-11635 Athens, Greece} 
\affiliation{Fachbereich Physik, Universit{\"a}t Konstanz, D-78457 Konstanz,
Germany}
\author{Santiago Cuesta-L\'opez}
\affiliation{ICCRAM, University of Burgos, Science and Technology Park, Plaza
Misael Ba{\~{n}}uelos, 09001 Burgos, Spain}
\author{Dimitar Angelov}
\affiliation{Universit{\'e} de Lyon,
Laboratoire de Biologie et de Mod{\'e}lisation de la Cellule (LBMC)
CNRS/ENSL/UCBL UMR 5239, Ecole Normale Sup{\'e}rieure de Lyon, 69007 Lyon,
France}
\author{Tobias Unruh}
\affiliation{Lehrstuhl f{\"u}r Kristallographie und Strukturphysik,
Friedrich-Alexander-Universit{\"a}t 
Erlangen-N{\"u}rnberg, Staudtstr. 3, D-91058 Erlangen,
Germany}
\author{Michel Peyrard}
\affiliation{Universit{\'e} de Lyon, Ecole Normale Sup{\'e}rieure de Lyon,
Laboratoire de Physique CNRS UMR 5672, 46 all{\'e}e d'Italie, F-69364 Lyon
Cedex 7, France}
\email[To whom correspondence should be addressed. ]
{Michel.Peyrard@ens-lyon.fr}

\begin{abstract}
DNA is a flexible molecule, but the degree of its flexibility is subject to
debate.  The commonly-accepted persistence 
length of $l_p \approx 500\,$\AA\ is
inconsistent with recent studies on short-chain DNA  that show
much greater flexibility but do not probe its origin.  
We have performed X-ray and neutron small-angle
scattering on a short DNA sequence containing a strong nucleosome positioning
element, and analyzed the results using a modified Kratky-Porod model to
determine possible conformations.  Our results support a hypothesis from Crick
and Klug in 1975 that some DNA sequences in solution can have sharp kinks,
potentially resolving the discrepancy.  Our conclusions are supported by
measurements on a radiation-damaged sample, where single-strand breaks lead to
increased flexibility and by an analysis of data from another sequence, which
does not have kinks, but where our method can detect a locally enhanced
flexibility due to an $AT$-domain.
\end{abstract}

\maketitle

\section{Introduction}
\label{sec:intro}

Remarkable progress has been achieved since the famous publication of the DNA
structure by Watson and Crick \cite{WATSON-CRICK} but basic questions are
still open. One of them was raised by Crick and Klug in 1975 \cite{CRICK-KLUG}
when they examined the ease with which duplex DNA can be deformed into a
compact structure like chromatin. They suggested that DNA can form a sharp
bend, that they called a kink. Kinks are not merely sharp bends due to
fluctuations or a broken strand, but metastable structures which can exist
without a drastic distortion of the
configuration of the backbone. They have been subsequently found in
highly-constrained environments, such as in chromatin, where other molecules
force sharp bends \cite{ONG,CHUA}.  However Crick and Klug raised a more
fundamental question: could DNA kinks occur spontaneously as a result of
thermal motion?

Since then this hypothesis was never confirmed and the origin ot the
flexibility of DNA is still debated \cite{KAHN}.  The molecule is often
described as a cylinder, or worm-like chain (WLC), that is relatively stiff
with a persistence length of $l_p \approx 500$ {\AA}
\cite{MARKOSIGGIA,YANMARKO,MASTROIANNI}.  
This value is at odds with the tight-packing of
DNA in chromatin and with recent experiments on short-chain DNA that show much
shorter persistence lengths \cite{CLOUTIER2004,CLOUTIER2005,YUAN,VAFABAKHSH}.  
However the
studies are somewhat limited in that they do not probe the full conformation
of the molecule and cannot identify the position or the degree of any
bending.  

\begin{figure}[h]
\begin{center}
\includegraphics[width=8cm]{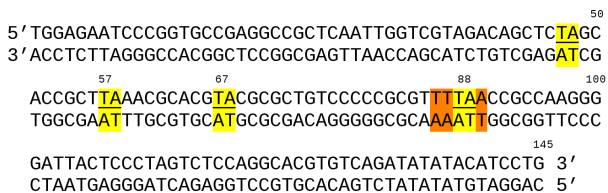}
\end{center}
\vspace{-0.5cm}
\caption{Sequence of the 145 base-pair DNA molecule
   investigated in this study. This sequence was used to build
   the NCP-601 nucleosome core particle investigated in
    \cite{VASUDEVAN}. The fragment with orange background is the strong
    positioning element, 
    characteristic of this sequence. The TA fragments considered as possible
    kink positions in the model are underlined and marked in yellow.
}
  \label{fig:seq}
\end{figure}

Small-angle scattering (SAS) experiments are excellent 
techniques to investigate particles
with linear dimensions $\sim 1 -100$ nm and are therefore well-suited to
studying short-chain DNA.  They probe the spatial distribution of the
scattering length density and are very sensitive to the overall shape and size
distribution of particles.  In the dilute limit, SAS probes the ensemble
average of all the orientations and shapes that the particles adopt over the
duration of the measurement.  We have carried out measurements with both
X-rays (SAXS) and neutrons (SANS) of short-chain DNA in solution, free from
any molecular construct 
like the addition of fluorophores \cite{VAFABAKHSH} or gold nanoparticles 
\cite{MATHEW-FENN}. The principal experiments were performed with
SAXS on a 145 base-pair sequence of DNA
containing the ``601'' strong positioning sequence \cite{LOWARY} known for
easy wrapping around a histone core.  The sequence, which has been
investigated in structural studies of nucleosome core particles
\cite{VASUDEVAN}, is shown in Fig.~\ref{fig:seq} and is henceforth called
``Widom-601''.  The results have been analyzed using a dynamical model.
This analysis shows the presence of kinks at
positions consistent with the hypothesis of Crick and Klug.
As a further test, a complementary experiment has been made with a different
DNA sequence, with 204 base pairs, derived from a piece of
$\lambda$-phage DNA, which has been  modified to introduce a segment
containing only 
$A-T$ pairs from sites 94 to 125. This introduces a domain with a higher local
flexibility and some weak intrinsic curvature due to intrinsically curved
elements (such as AATT). The same analysis, performed on the data collected
with this control sequence did not detect kinks but the statistics of the
local bending angles showed a increased average value, and increased
fluctuations, in the domain where we introduced the AT rich segment,
showing the ability of our analysis to detect detailed features of DNA
flexibility with SAS.

\section{Samples and experimental methods}
\label{sec:matmed}

\subsection{Sample preparation}

The main sample of this study is the ``Widom-601''
145 base-pair sequence of DNA shown in Fig.~\ref{fig:seq}. It
contains the ``601'' strong positioning sequence \cite{LOWARY}, known for
easy wrapping around a histone core.
Multiple repeats of this nucleic sequence
were inserted into the EcoRV site of the pGEMT
easy vector and expressed in E.Coli DH5$\alpha$ cells. The fragments were
excised from the vector by restriction enzyme EcoRV, 
followed by phenol chloroform
extraction and ethanol precipitation. The excised 145bp DNA were separated
from the linearized plasmid through 5{\%} polyacrylamide gel 
electrophoresis (prep
cell-BioRad) and the purity of the sample was analyzed by 1{\%} agarose gel
electrophoresis and PAGE. This preparation method provides 100\% homogeneous
DNA of crystallographic-grade purity and was also applied to prepare
samples for  nucleosome crystallization, used in other 
experiments \cite{BEDNAR-ANGELOV}. 

\medskip
\begin{figure}[h]
\vspace{-0.3cm}
  \centering
  \includegraphics[width=7cm]{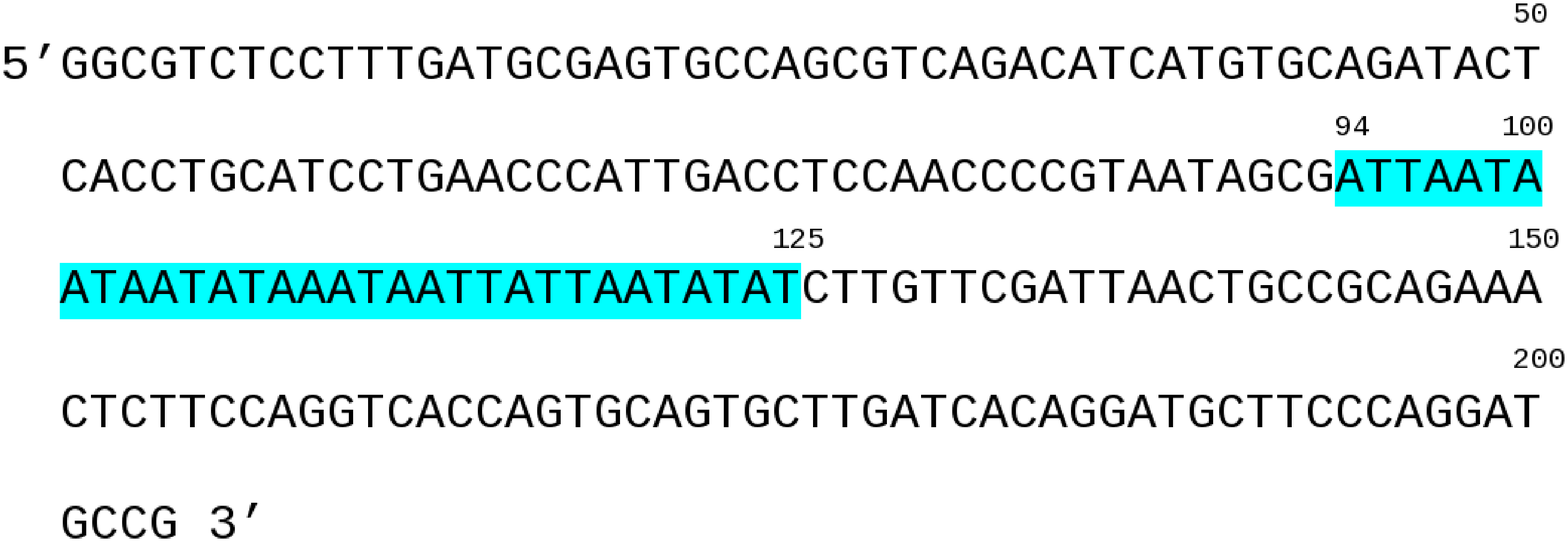}
\vspace{-0.3cm}
  \caption{Sequence of 204 base pairs studied in a complementary
    experiment. In the 
    fragment marked in blue the original sequence part of $\lambda$-phage DNA
  has been modified. The $G-C$ pairs have been changed to $A-T$ pairs to
  create a domain of 32 consecutive $A-T$ pairs.}
  \label{fig:seqsans1}
\end{figure}

The same method was used to prepare the modified $\lambda$-phage sequence
shown in Fig.~\ref{fig:seqsans1},
studied in complementary experiments presented in
Sec.~\ref{sec:complementary}. 

Prior to the SAXS and SANS experiments the samples were dissolved in a
$^2$H$_2$O buffer containing 10mM Tris (pH 7.4), 0.1 mM EDTA and 30 mM NaCl
with a DNA concentration of 2.2 mg/mL for the Widom-601 sequence and 1.27
mg/mL for the modified $\lambda$-phage sequence, and degassed in a partial
vacuum of 0.5 bar for 3 hours before being loaded into quartz containers for
measurement.  The quartz cells for SANS were rectangular with a 2 mm
thickness.  Quartz cylindrical capillaries with an inner diameter of 1 mm and
a wall thickness of $10\;\mu\mathrm{m}$ were used for SAXS.

\subsection{Small angle scattering}
Small angle scattering of neutrons and X-rays by dilute particles in solution
probes the scattering length density of the particles. 
The scattering cross-section is given by \cite{SIVIA}:

\begin{equation}
	\label{eq:SASxsec} \frac{\text{d}\sigma}{\text{d}\Omega} \propto
\frac{N}{V} \left<\left| \iiint\limits_V
\beta\left(\mathbf{r},\bf{\Theta}\right)
\exp\left({i\mathbf{q}\cdot\mathbf{r}}\right)\text{d}^3\mathbf{r}\right|^2
\right>_{\bf{\Theta}},
\end{equation}
where the sample contains $N$ particles in volume $V$, and $\mathbf{q}$ is the
momentum transfer.  The scattering length density at position $\mathbf{r}$ in
a particle is given by $\beta\left(\mathbf{r}\right) =
\rho\left(\mathbf{r}\right)\overline{b}\left(\mathbf{r}\right)$, with
$\rho\left(\mathbf{r}\right)$ being the local atomic number density.  
The
local mean scattering length, $\overline{b}\left(\mathbf{r}\right)$, varies as
a function of atom and isotope for neutrons \cite{SEARS} and as the Thomson
scattering length multiplied by the local number of electron for X-rays.  The
triangular brackets in equation (\ref{eq:SASxsec}) indicate that the
cross-section measures an ensemble average of all the orientations,
$\bf{\Theta}$, and configurations that the particles have in the sample.

\subsubsection{Small angle X-ray scattering (SAXS)}
The SAXS measurements were performed using the VAXTER instrument at the
Friedrich-Alexander Universit{\"a}t, Germany.  It uses Ga-K$\alpha$1,2
radiation ($\lambda =1.34\;$\AA) from a GaMetalJet D2 70 kV X-ray source
(EXCILLUM, Kista, Sweden) with 150 mm Montel optics (INCOATEC, Geesthacht,
Germany).  X-ray-absorbing diaphragms defined the incident beam collimation.
The sample-detector distance was set to 1.5 m, and the instrument uses a
Pilatus3 $300\;$K detector.  The sample temperature was controlled by an
external water bath.  The data were corrected for transmission and background.
The scattering from a glassy carbon standard \cite{ZHANG2010} was used to
normalize the data to absolute scale.  The resulting cross-sections covered the
range $0.006 \leq Q \leq 0.2$ {\AA}$^{-1}$.

\subsubsection{Small angle neutron scattering (SANS)}
The SANS measurements were performed using the D22 instrument at the Institut
Laue-Langevin, France.  The incident beam wavelength was set to $\lambda = 6
${\AA} with $\Delta\lambda/\lambda =0.1$.  The incident beam collimation was
defined using neutron-absorbing diaphragms.  The sample temperature was
controlled by an external water bath.  Data were recorded with the
sample-detector distance at 5.6 and 17.6 m, 
with the collimation set at 5.6 m and
17.6 m respectively for the optimal compromise between resolution and beam
flux.  The GRASP data reduction suite 
(https://www.ill.eu/instruments-support/instruments-groups/groups/lss/grasp/home/)
was used to correct the
data for instrument background, empty cell, detector efficiency and to
normalize by the direct beam intensity to obtain the intensity in 
absolute scale.  The data
were then merged, buffer subtracted, and radially integrated to give the
cross-section in the 
range $0.003 \leq Q \leq 0.15$ {\AA}$^{-1}$.

\begin{figure*}
\begin{center}
\begin{tabular}{ll}
\includegraphics[height=5.4cm,clip]{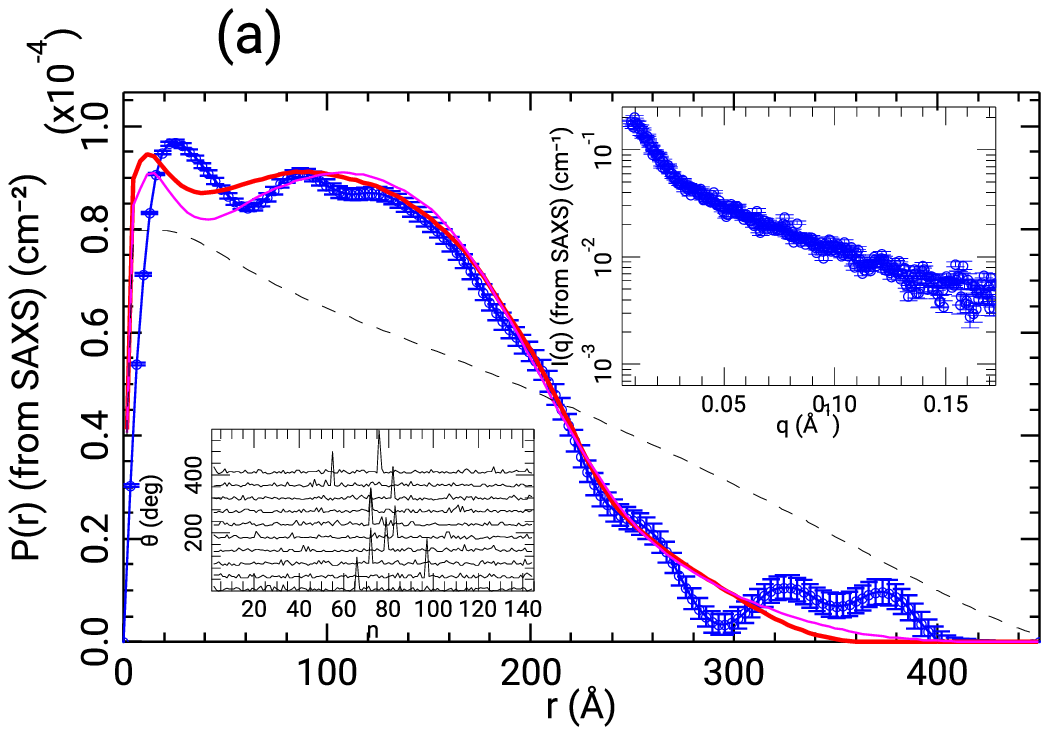} &
\includegraphics[height=4.7cm,clip]{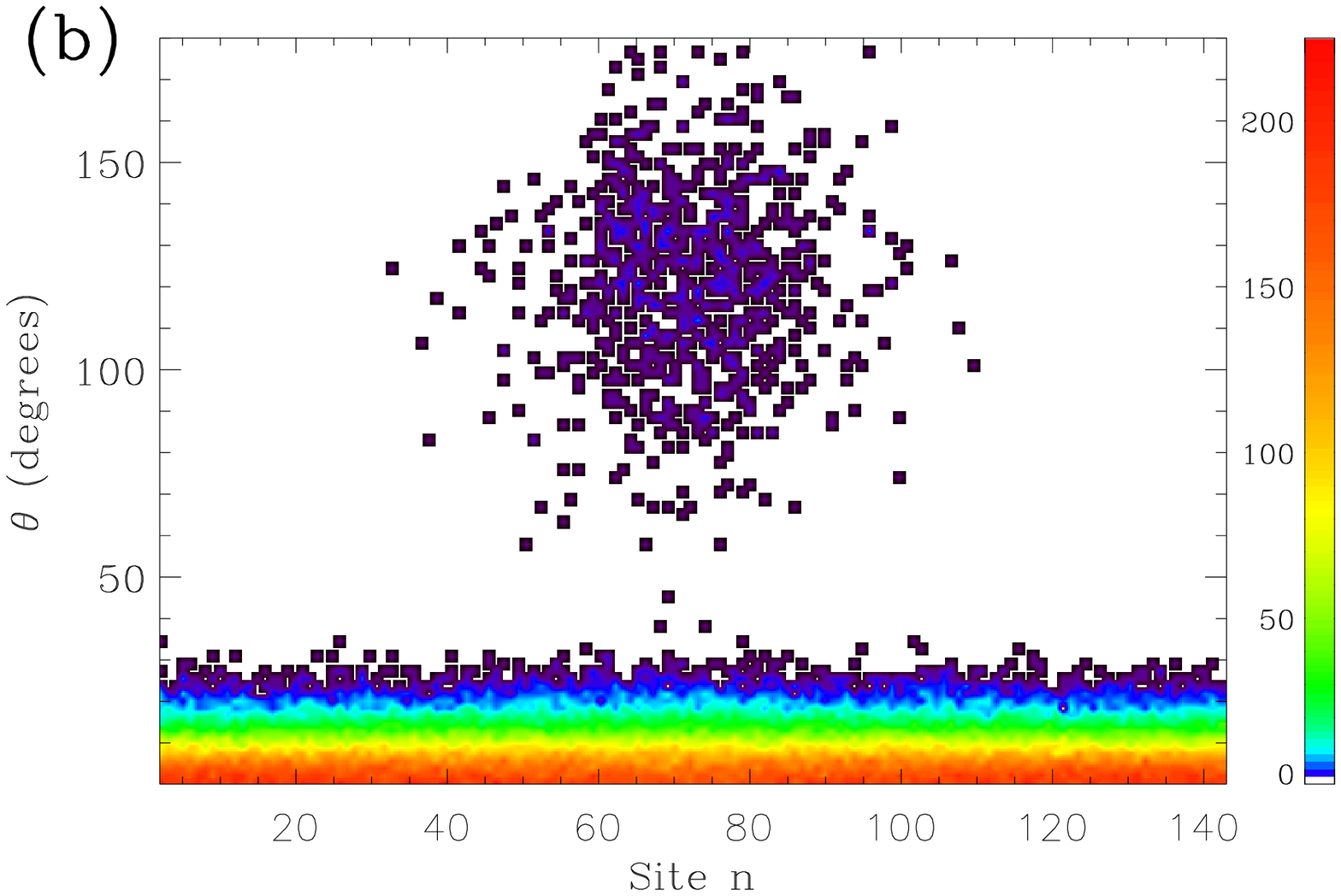}
\hspace{-0.3cm}
\includegraphics[height=4.5cm,clip]{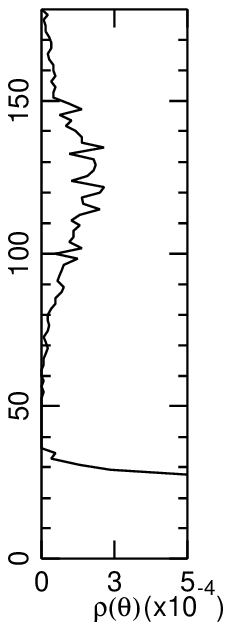}\\
\includegraphics[height=5.4cm,clip]{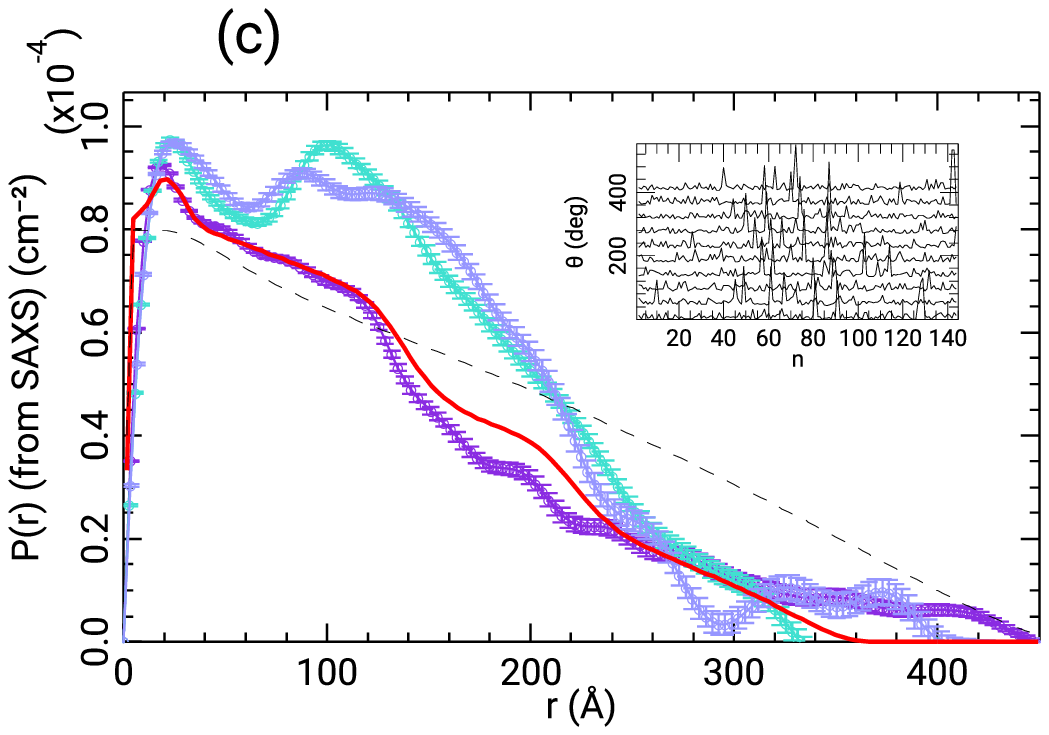} &
\includegraphics[height=4.7cm,clip]{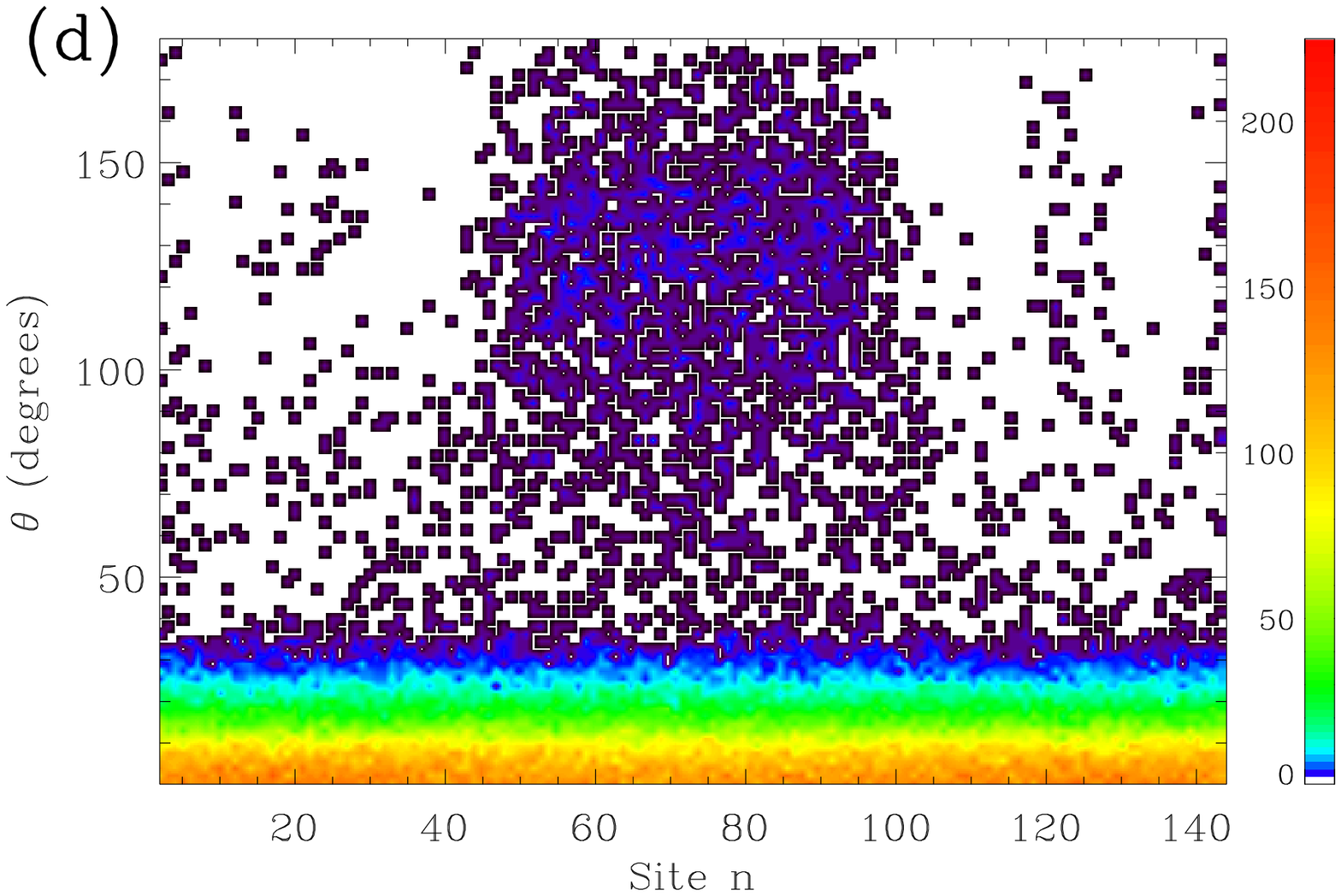}
\hspace{-0.3cm}
\includegraphics[height=4.5cm,clip]{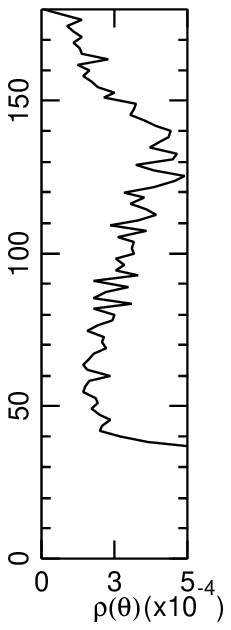} \\
\end{tabular}
\end{center}
\vspace{-0.5cm}
\caption{\setstretch{1.0}
  Experimental results and data analysis for the Widom-601 sequence.
(a) Blue: results of the SAXS experiment at room temperature. The top
right inset
shows the scattered intensity $I(q)$ versus the scattering vector $q$, and the
main panel shows the corresponding $P\left(r\right)$ computed by {\tt GNOM}
with estimated error bars. 
The thick red line shows the average $P\left(r\right)$
for the 10$^3$
conformations with the best match to the experimental data
among $12\times10^7$ conformations generated by Monte-Carlo.
Thin magenta line: $P\left(r\right)$ deduced from a Monte-Carlo simulation at
room temperature for the models with a kink of angle $80^{\circ}$ at site 67.
The dashed black line shows $P\left(r\right)$ for a homogeneous
Kratky-Porod model having a persistence length of $500\,$\AA\
at room temperature ($K = 150$, $\theta_{0,} = 0$, $C = 0$).
The bottom left inset shows the values of the local bending angles for the 10
model conformations which provide the best match with the experimental 
$P\left(r\right)$. Successive plots are moved up by $50^{\circ}$ to limit the
overlap between the curves.
(b) Histogram of the bending angle
against $n$ for the 10$^3$ conformations that provide the best matching with
the room temperature SAXS data.
For each site of the polymer model, the left part shows the number of
 $\theta_n$ values that correspond to the value marked on the left scale (the
 total of these numbers for a given $n$ is equal to 1000, the number of
 conformations) with a color scale shown on the right. The right part of the
 figure shows the fraction $\rho$ of $\theta$ angles, integrated over the whole
 model, which belongs to a given range of theta. The scale is truncated to
$\rho_{\mathrm{max}} = 5 \times 10^{-4}$ to better show the part of the curve which
corresponds to large $\theta$ angles.
(c) Dark-blue: The SAXS $P\left(r\right)$ at $70^{\circ}$C
after a long exposure
to X-rays. The thick red curve shows the average  $P\left(r\right)$ for
the 1000 conformations that provide the best match with the experimental data
has been obtained from a search of $70\times10^7$ conformations generated by
Monte-Carlo. Light blue: The SAXS $P\left(r\right)$ at room temperature
(also shown in blue in panel (a)) for comparison. 
Turquoise: The SAXS $P\left(r\right)$
 at $70^{\circ}$C with short exposure to X-rays.
The inset shows the values of the local bending angles for the 10
model conformations which provide the best match with the experimental 
$P\left(r\right)$ as in panel (a).
(d) same as panel (b) for the 10$^3$ conformations that provide the best
matching with long-exposure SAXS data. The color scale and the scale for
$\rho$ in the right part of the plot are the same as for panel (b) to allow a
quantitative comparison between the two cases.}
\label{fig:summary}
\end{figure*}

\section{Experimental results on the Widom-601 sequence}
\label{sec:expwidom}

The SAS scattering cross-section for dilute particles in solution is given by
the one-dimensional integral:
\begin{equation}
	\label{eq:SASxsec1D} \frac{\text{d}\sigma}{\text{d}\Omega} \propto
\int\limits_0^{d_{max}} P\left(r\right)
\frac{\sin\left(qr\right)}{qr}\text{d}r,
\end{equation}
where $q$ is the momentum transfer and $d_{max}$ is the longest dimension in
the object.   The pair-distribution function, $P\left(r\right)$, correlates
the scattering length densities for volume elements separated by a distance
$r$ within the particle and is further weighted by the contrast between the
scattering length densities of the particle and the solution.  The
$P\left(r\right)$ therefore differs for X-rays and neutrons, however, for
scales beyond about $40\;$\AA\ which do not resolve the internal structure of
DNA and are only sensitive to the conformation of the molecular axis, both
techniques are expected to bring the same information.

Mathematically,
$P\left(r\right)$ is obtained from an inverse Fourier transform of the
cross-section.  We used the {\tt GNOM} program \cite{SVERGUN} with manually
set parameters to calculate $P\left(r\right)$. 
The $P\left(r\right)$ derived from the SAXS data at $23^{\circ}$C is shown
Fig.~\ref{fig:summary}-a.  The measurement took one hour.
Temperature-dependent $P\left(r\right)$ derived from the SAXS data are shown
in Fig.~\ref{fig:summary}-c.  Two samples were measured at high temperature.
For the first, labeled ``short exposure'',  the sample was heated directly to
$70^{\circ}$C and measured for one hour.  For the second, labeled ``long
exposure'', the data result from the final measurement of a sequence of
hour-long runs at $30^{\circ}$C, $50^{\circ}$C and $70^{\circ}$C.

The ``short exposure'' $P\left(r\right)$ is very similar to that at
$23^{\circ}$C, while the ``long exposure'' data show dramatic differences due
to X-ray radiation damage.  At low X-ray flux, most of the damage occurs as
single-strand breaks due to free radicals created by the interaction of X-rays
with the water molecules \cite{VONSONNTAG, ABOLFATH}. This is likely to be
exacerbated at high temperatures. In our SAXS experiment about 50\% of the
X-rays were absorbed by the solvent, and the influence of radiation damage is
clearly visible.

\bigskip
Prior to the SAXS measurements
preliminary measurements on the Widom-601 sequence
had been performed with 
SANS as a function of temperature up to
$79^{\circ}$C. 
\emph{In-situ} UV absorption spectroscopy recorded during the SANS
measurements showed that there was no significant thermal denaturation of
double-stranded DNA up to this temperature.

\begin{figure}[h]
\vspace{-0.3cm}
  \centering
  \includegraphics[width=8cm]{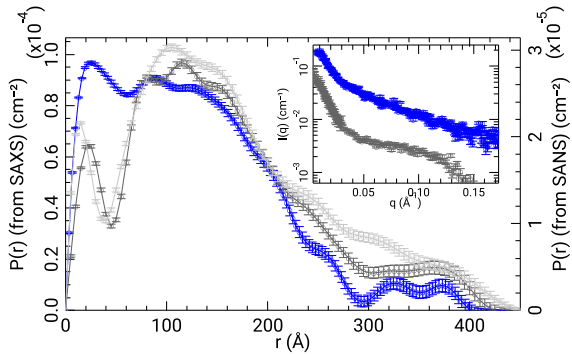}
  \caption{Comparison of the SAXS and SANS results for the Widom-601 sequence.
The SAXS data are shown in blue and the neutron data are shown in grey (dark
grey: measurements at $15^{\circ}$C, light grey: measurements at
$79^{\circ}$C. The inset shows the scattered intensity versus the scattering
vector $q$. The corresponding $P(r)$ have been obtained by GNOM. Their
normalization is such that $\int P(r) dr = I(q=0)$.}
  \label{fig:comparexn}
\end{figure}

The SANS measurements up to $79^{\circ}$C did show some
limited changes versus temperature (Fig.~\ref{fig:comparexn})
which could come from an increase of the thermal
fluctuations, but the overall shape of $P\left(r\right)$ was only weakly
modified.  Radiation damage from neutrons was expected to be negligible as
they are non-ionizing, and the SANS data showed no strong change with exposure
time.

Figure
\ref{fig:comparexn} compares the SAXS and SANS results.
The $P\left(r\right)$ show common features. However there are differences
because SAXS is more sensitive to heavier elements as $\rho\overline{b}$ is
proportional to the electron density for X-rays.  The SAXS $P\left(r\right)$
is thus dominated by the correlations between the phosphates.  The neutron
$P\left(r\right)$ is more sensitive to the distribution of hydrogen in the
sample. Both data sets show a peak at $r \approx 20$ {\AA}.  The peak in the
X-ray data is due to the phosphates, which are on the outside of the double
helix and are 20 {\AA} apart.

Due to the many protons distributed on various sites in the
  DNA structure, for short distances neutron scattering is well approximated
  by the scattering by a bulk cylinder with a diameter of $20\;$\AA, which
  also shows a peak in $P(r)$ in the vicinity of the diameter of the cylinder.
Therefore, both for SANS and SAXS, the
$20\;$\AA\ peak appears to be signature of the diameter of the double helix.

For larger $r$ distances the differences between SAXS and SANS are expected to
decrease because at this scale the differences between the atom types are no
longer resolved and $P\left(r\right)$ essentially reflects to conformation of
the molecules. 
For $r \ge 70\;$\AA, two curves for $P\left(r\right)$ have a similar
shape. They show a roughly flat region in the $70-160\,$\AA\ range and then
decrease strongly before showing a kind of plateau in the
$300 - 400\;$\AA\ range.

\section{Data analysis: Conformation of the Widom-601 DNA molecules in
  solution}
\label{sec:anawidom}

We focused our analysis of the Widom-601 data on the SAXS data which provided
a stronger and less noisy signal than the SANS experiments.

\subsection{Standard SAS data analysis}

As a first step we relied on standard SAS data analysis packages.
The $23^{\circ}$C SAXS data were analyzed using the 
SASVIEW software \cite{SASVIEW}
to model DNA as an homogeneous flexible cylinder, giving a persistence length
of $l_p = 97.9$ {\AA}.  A further analysis used the Kratky-Porod model
\cite{KRATKY} which is the discrete version of the worm-like chain model
generally used for long DNA molecules \cite{MARKOSIGGIA,YANMARKO} and whose
structure factor can be calculated exactly \cite{NTH-KP}.  This calculation
gave a best fit of $l_p = 117$ {\AA}.  Both values are much smaller than
measured values for double-stranded DNA \cite{KAHN,HAGERMAN}, including the
generally-accepted  value of $l_p \approx 500$ {\AA}.  They are also
unrealistically small considering that the double helix has a diameter of
$20\;$\AA.  The analysis confirms the conclusions, drawn from accurate
measurements on short DNA molecules \cite{VAFABAKHSH}, that a 
homogeneous WLC model does not describe the flexibility of short-chain DNA. 

However, it is important to stress that the persistence length obtained in
these fits is actually an effective persistence length which includes both the
effects of the statistical fluctuations and a contribution of the intrinsic
curvature of the molecule. As pointed out by Schellmann and Harvey
\cite{SCHELLMAN}, any kind of bend, stiff or flexible, will result in a reduced
persistence length and, as a result, permanent structural kinks can reduce the
persistence length.

\bigskip
\begin{figure}[h]
\begin{center}
  \begin{tabular}{cc}
    \includegraphics[height=4.8cm]{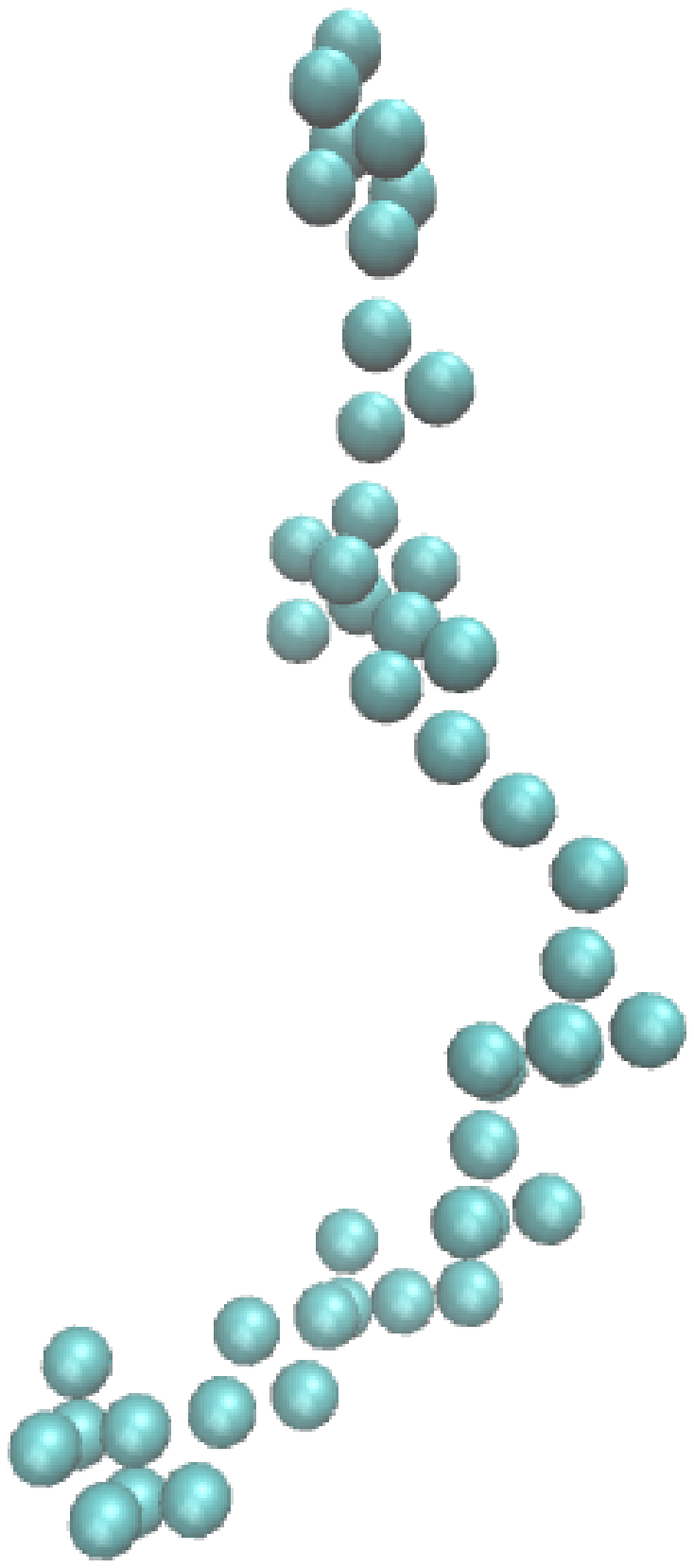} &
    \includegraphics[height=4.8cm]{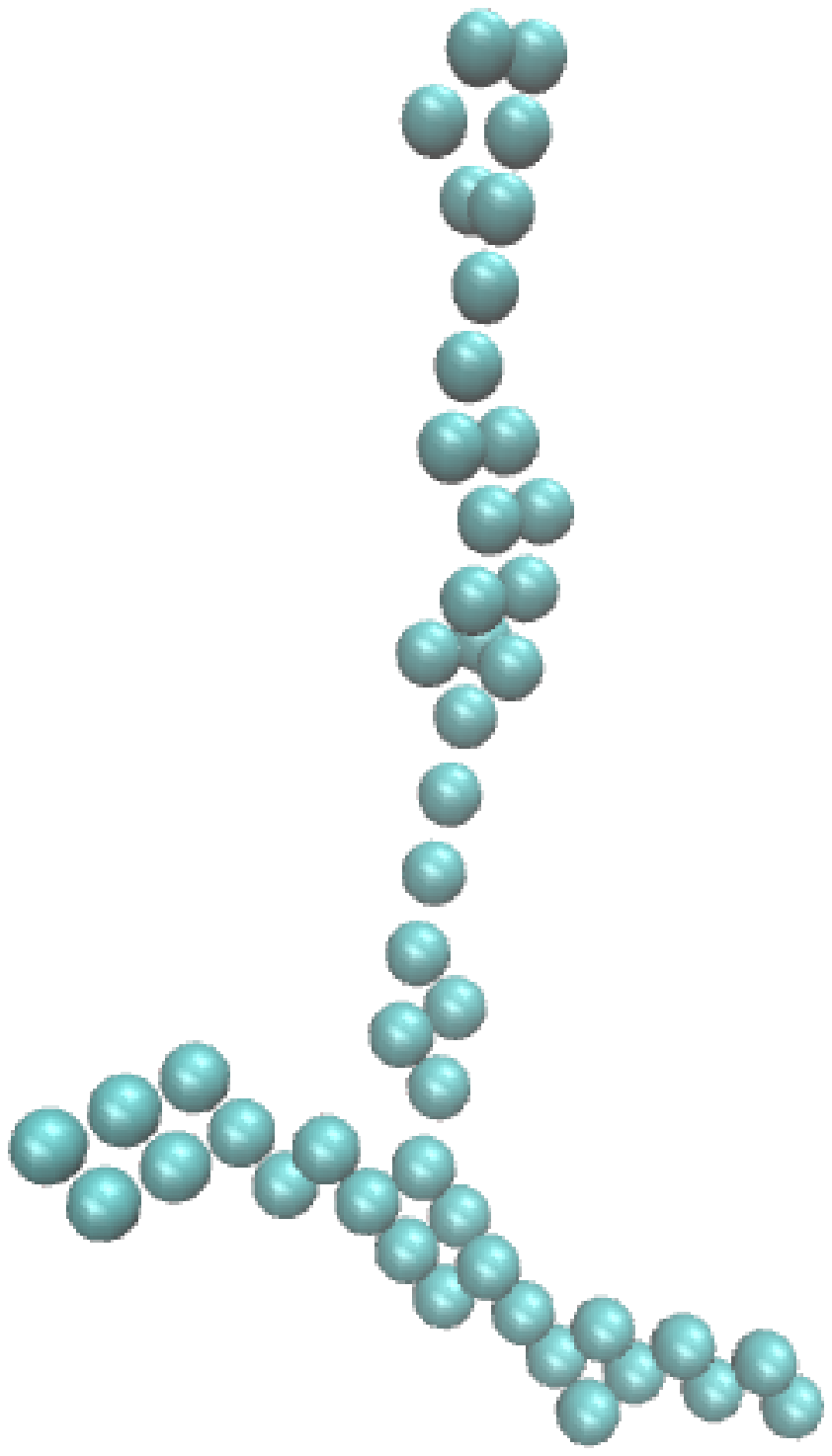}
  \end{tabular}
\end{center}
\vspace{-0.5cm}
\caption{Examples of the output of the {\tt DAMMIF} program, showing possible
  optimized shapes of the DNA molecules in solution}
  \label{fig:dammif}
\end{figure}

Therefore, to understand what appears as an anomalously low persistence length,
the next step was to attempt to reconstruct the
main features of the particle shapes from the data. However this is not
straightforward because the knowledge of
$P\left(r\right)$ is not sufficient to unambiguously determine 
this shape, even for the case of a linear polymer.  
Iterative schemes, starting from a broad and random conformational search and
then progressively improving by testing likely models, have been developed.
Within the ATSAS package \cite{PETOUKHOV}, the {\tt DAMMIF} program 
\cite{SVERGUN99,FRANKE} 
starts from the $P(r)$ given by the {\tt GNOM} program and uses
simulated annealing to optimize the shape of a set of dummy atoms
to retrieve the shape of the scattering molecules. 
The  program starts from a random configuration. A polymer like
DNA can have a broad variety of conformations in solution and multiple runs
lead to different final shapes. Nevertheless the results give hints
on the DNA conformations that fit the scattering data. When it is
applied to our SAXS and SANS data for the Widom-601 sequence, 
this program converges either to strongly curved
structures, or to branched solutions, as shown in Fig.~\ref{fig:dammif}.
As partial denaturation of our sample is ruled out by the UV measurements, we
do not expect any branching. However, if the solution actually contains a
mixture of weakly curved molecules while others have a sharp bend, these {\tt
DAMMIF} solutions could describe a superposition of two conformations
because $P\left(r\right)$ represents the ensemble average over all
configurations 
in the sample, including time-dependent fluctuations.

\subsection{Conformational search using a polymer model}

\begin{figure}
  \includegraphics[width=8.3cm]{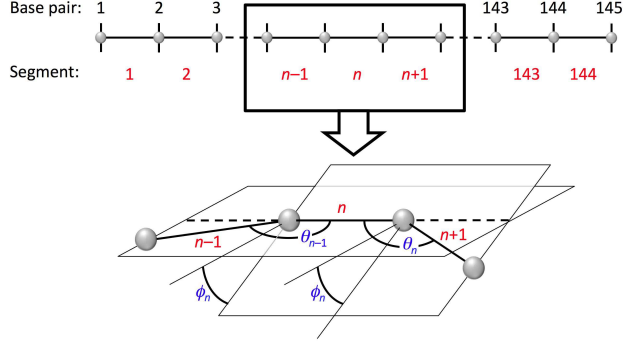}
\caption{Schematic showing the polymer model used for the WIDOM-601 DNA
measured in the experiments.  The top part of the figure shows the definition
for the numbering and the lower part shows the definition of the bond and
torsional rotation angles.}
\label{fig:ModelSchem}
\end{figure}

The {\tt DAMMIF} results show that fluctuations must be accounted for in
the modeling of the data.  Moreover the ab-initio shape reconstruction
performed by {\tt DAMMIF} does not include any {\em a priori} knowledge of the
molecular properties. It is more efficient to search among a set of
conformations derived from a model which take into account some of the known
features of the molecules.  However, to study the flexibility of the DNA
molecules, a model at the atomic scale or a coarse-grain model describing the
bases would be inefficient. Instead we describe DNA with an extension of the
Kratky-Porod model \cite{KRATKY}, i.e.\ the model, shown schematically on
Fig.~\ref{fig:ModelSchem}, is not concerned with the internal structure of the
DNA, but only with the conformations that the backbones could adopt.  It
consists of $N+1$ objects representing base pairs, each separated by $a =
3.34$ {\AA}, corresponding to the base-pair distance in DNA, with bond angles
that may vary at each point.  The model therefore has $N$ bonds, or segments.
The local angle at site $n$, which connects segments $n$ and $n+1$, is defined
as $\theta_n$, and the dihedral angle of rotation between the plane containing
segments $n-1$ and $n$ and the plane containing segments $n$ and $n+1$ is
defined as $\phi_n$.

Thermal fluctuations were accounted for by calculating the bending and
torsional energy for the model expressed through its Hamiltonian. A possible
permanent curvature is permitted by considering a local equilibrium value at
each site, given by $\theta_{0,n}$ and $\phi_{0,n}$, which may be different
from zero.  The Hamiltonian is thus given by:
\begin{align}
  \label{eq:hamil}
H = &\sum_{n=1}^{N-1} K_n [1 - \cos(\theta_n - \theta_{0,n})] \\ \nonumber &+
\sum_{n=2}^{N-1} C_n  [1 - \cos(\phi_n - \phi_{0,n})] \; ,
\end{align}
where $K_n$ and $C_n$ are constants setting the scale of the bending and
dihedral energies respectively. Dimensionless variables were used in the
calculations. Distances were measured in units of the base-pair distance $a$,
and temperature was measured in energy units and the energy was expressed
relative to $k_B T$ at room temperature, where $k_B$ is the Boltzmann
constant, so that the properties of DNA at room temperature were obtained by
setting $T=1$.  Equation \ref{eq:hamil}
reduces to the Kratky-Porod model \cite{KRATKY} on setting $C_n=0$ and $K_n=K$
for all $n$.  Setting $K = 150$ gives $l_p = 500$ {\AA}, while $K=35$ gives
the previously-discussed best fit to the $23^{\circ}$C data with $l_p = 117$
{\AA}.

For a given model conformation, $P\left(r\right)$ was calculated by
putting a unit scattering center at each base-pair position and then the
calculated $P\left(r\right)$ was scaled to have the same integrated area as
the data.  This gives a reasonable approximation for the SAXS data for $r
\gtrsim 40\;$\AA, as the X-ray scattering from a base pair is dominated by its
two phosphorus atoms whose contribution can be effectively mapped onto the
center of the base pair for large $r$. For small $r$ this evaluation of
$P\left(r\right)$ is a worse approximation for the SANS data which are far
more sensitive to the protons.

\bigskip
The model with $N = 144$, representing the Widom-601 sequence, was
used for a broad search through the conformational space to determine the
conformations which provide the best match with the experients.
At this stage of the analysis, our goal is not to design
a model of the DNA molecules in solution, but to extract information on the
molecular conformations from the data. {\em Therefore it is important to avoid
  any bias in the conformational search,} which could be introduced
for instance by assuming some {\em a-priori} knowledge such as
sequence dependent bending angles.
The bending constant was
set to $K_n=K=150$.  The torsional constant was set to $C_n=C=2$ which leads
to a clear dominance of $\phi \approx 0$, in agreement with the torsional
rigidity of DNA, but nevertheless allows large fluctuations.   The equilibrium
angles were set to $\theta_{0,n} = \phi_{0,n} = 0$.

Up to $12 \times 10^7$
Monte Carlo-generated conformations were created and accepted with the
probability $\exp (- H / T)$ at a temperature of $T = 3$, i.e. three times
room temperature, to widely explore the conformational space.   The
$P\left(r\right)$ were calculated for accepted conformations and compared to
the SAXS $P\left(r\right)$ at 23$^{\circ}$C by computing the standard
deviations between the two functions. The best $10^3$ conformations were
selected. They can be considered as representative of the conformations 
of the DNA molecules in solution so that their analysis allowed us to
determine the main features of the those conformations.
Their average $P\left(r\right)$  is plotted in
Fig.~\ref{fig:summary}-a.  The agreement with the experimental
$P\left(r\right)$ is satisfactory, and is excellent in the region 80 {\AA}
$\leq r \leq 280$ {\AA}. This is the critical range to assess the shape of DNA
molecules, which have a length of $484\;$\AA\ and a radius of gyration of
about $100-150\;$\AA.  A calculation of the WLC with $l_p = 500$ {\AA} is also
shown in Fig.~\ref{fig:summary}-a for comparison.  It does not resemble the
experimental $P\left(r\right)$. 

Fig.~\ref{fig:summary}-b shows a histogram of $\theta_n$ at the various sites
for the best $10^3$ conformations. The vast majority of $\theta_n$ are less
than 30$^{\circ}$, consistent with thermal fluctuations around $\theta_{0,n} =
0$.  However there is a significant concentration of large $\theta_n$ near the
center of the molecule.  The distribution is centered at $\theta_n \approx
125^{\circ}$, and it is distinct from the distribution at lower angles.
Inspection of those conformations with large $\theta_n$ revealed that each had
one, and only one, of the sites with a large angle.  The insert on
Fig.~\ref{fig:summary}-a shows the local bending angles for the 10 best
conformations.  The modeling suggests that the Widom-601 sequence has
metastable states with large angles in its central region, consistent with the
kinks proposed by Crick and Klug.  Kinks would also help explain the branched
conformations given by the {\tt DAMMIF} program if the solution contains a
mixture of kinked and non-kinked DNA molecules.

\subsection{Test of a model for kinked DNA molecules in solution}

As a blind conformational seach hints that some conformations may have a
\emph{permanent kink}, a second stage of our analysis was to check this
hypothesis. In this stage the polymer model is used in a different
context. Instead of chosing a generic set of parameters to allow an unbiased
scan of possible conformations, we use the results of the conformational
search to select specific parameters, and then run a Monte Carlo
 at room temperature ($T = 1$ in our dimensionless units) to test the
agreement of the resulting $P(r)$ with the experimental results. The possible
presence of a kink is tested by  
setting $\theta_{0,n} = \theta_\kappa$ for specific $n = n_\kappa$.
Following the hypothesis of Crick and Klug, we
identified A$-$T sites in the central region and explored conformations with
$70^{\circ} \leq \theta_{0,n} = \theta_\kappa \leq 110^{\circ}$ at these
sites.  Inspection
of Fig.~\ref{fig:seq} shows A$-$T sites at $n_\kappa = 47$, 57, 67, and 88.
The last site is of particular interest because it is part of the strong
histone-positioning element characteristic of this sequence.  The values for
$K_n$, $C_n$, $\phi_{0,n}$ and $\theta_{0,n}$ were all maintained as for the
conformational search, except for the one $\theta_\kappa$.

\medskip
To quantitatively evaluate the different possibilities, we calculated 
 \begin{equation}
  \label{eq:chi2}
  \chi^2 = \frac{1}{N_r - 1}\sum_1^{N_r} 
 \left[\frac{P_{\mathrm{theo}}(r_j) - P_{\mathrm{exp}}(r_j)}{\sigma_j}\right]^2
\end{equation}
where $r_j$ are the points where $P_{\mathrm{exp}}\left(r\right)$ has been
computed, $\sigma_j$ the experimental error at these points (given by the {\tt
GNOM} program) and $N_r$ the number of calculation points. Instead of the
simpler standard deviation used for the conformational search, which is
equivalent to setting $\sigma_j=1$ in Eq.~(\ref{eq:chi2}), here we take into
account the actual experimental errors to measure the validity of the model
more accurately. The summation is
restricted to $r_j > 50\,$\AA\ because the polymer model, which does not take
into account the diameter of the molecules, cannot be expected to properly
describe the properties of the DNA molecules at very small distances. 

\begin{table}[ht]
\medskip
  \centering
  \begin{tabular}{ccccc}
\hline
$n_K$ & $\theta_K$ (deg) & $\chi^2$ kink \\
\hline
88       & 95    & 2.31 \\
88       & 110   & 9.17 \\
88       & 80    & 1.26 \\
88       & 70    & 3.00 \\
67       &95     & 3.32 \\
67       & 110   & 9.17 \\
\textbf{67}       & \textbf{80}    & \textbf{0.88} \\
57       & 95    & 2.14 \\
47       & 95    & 2.91 \\
\hline
  \end{tabular}
  \caption{Results of Monte Carlo simulations at room temperature ($T=1$ in
    reduced units) for various kink positions $n_K$ and angles
    $\theta_K$. The table lists the values of the $\chi^2$ distance between
    the experimental and theoretical $P\left(r\right)$ (Eq.~(\ref{eq:chi2})) 
for kinked DNA (3rd column). The line in bold-face is the case which provides
the 
best agreement with experimental data. For non-kinked DNA, the $\chi^2$
distance between the theoretical and experimental probability distributions
(Eq.~(\ref{eq:chi2})) is $\chi^2 = 22.86$.}
  \label{tab:resuchi2}
\end{table}

\begin{figure}[h!]
\vspace{-0.2cm}
\begin{center}
\includegraphics[width=7.0cm]{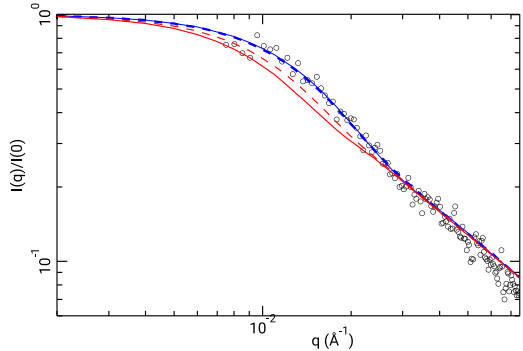}
\end{center}
\vspace{-0.8cm}
      \caption{comparison between the SAXS structure factor (circles) 
       and the structure
        factor of the polymer model, at room temperatures, for different
        parameters: i) thin blue curve, model with a kink of $80^{\circ}$ at
        position $67$, ii) dashed blue curve, almost identical to the previous
        one, kink of $80^{\circ}$ at
        position $88$, iii) full red curve, Kratky-Porod model, 
        without kink and a
        persistence length of $500\;$\AA, iv) dashed red curve, Kratky-Porod
        model, without kink and a 
        persistence length of $224\;$\AA  }
\label{fig:sqK}
\end{figure}

Figure \ref{fig:sqK} compares the SAXS structure factor
with the structure factor of the polymer model with
a kink at one of the two $TA$ steps closest to the center, $n_K = 88$ (thin
blue curve) or $n_K=67$ (dashed blue curve) and with the structure factor of a
homogeneous WLC model, for two values of the persistence length $l_p =
500\;$\AA\ and $l_p = 224\;$\AA.

The results for different kink positions and angles are given in Table
\ref{tab:resuchi2}. Although such a simple polymer model cannot
describe all the fine structure of the experimental results, in the range
where the model applies the theoretical results fit the data
within the experimental
error bars in the range $50-300\;$\AA. For $r < 50\,$\AA\ the model should be
completed by taking into 
account the diameter of the DNA molecule and its internal structure, while
for $r > 300\,$\AA\ some extra flexibility near the ends might be necessary
to take into account the fluctuations of the twist at the free boundaries.

Table \ref{tab:resuchi2} shows that the best agreement with experiments is
obtained for $n_K = 67$, although the position $\theta = 88$, within the
$TTTAA$ strongly positioning sequence, gives results which are almost equally
good. Our analysis cannot rule out one of the two optimal kink sites. The
case of site $n_K=57$, which also gives good results is 
almost symmetric of $n_K=88$ with respect to the center of the molecule. For
molecules in solution, at the scale of the SAXS/SANS experiments, 
and for the model, the two ends are indistinguishable
and the $57$ and $88$ sites are therefore almost equivalent. 
Moreover, looking at the sequence of the two strands listed in
Fig. 1 of the article,
one can notice that, around site 57 one finds
that same sequence of 5 base pairs as around site 88.
The best agreement with experiments is
obtained for a kink angle of $80^{\circ}$, in good agreement with the kink
predicted by Crick an Klug \cite{CRICK-KLUG}.

The tests were not exhaustive, but they establish that models with
a kink in the DNA give far better comparisons with the experimental results
than a WLC model with a sensible $l_p$.  

\subsection{Analysis of the long-exposure SAXS data}

The results pointing to the possible existence of kinks for the Widom-601
sequence in solution are supported by the analysis of the
SAXS data at high temperature.  A
conformation search with all equilibrium angles equal to zero was applied  to
the ``long exposure'' data.  The search was performed at higher temperature of
$T = 6$ to allow for an even broader range of conformations.  More than $7
\times 10^8$ conformations were generated in the Monte Carlo search, and the
average $P\left(r\right)$ from the best $10^3$ conformations are shown in
Fig.\ref{fig:summary}-c with the histogram of $\theta_n$ vs. $n$ shown in
Fig.\ref{fig:summary}-d.  The results show many sites with large $\theta_n$
distributed more evenly across the sequence and over intermediate angles. 
The distribution in $\theta_n$ is also broader, with no gap between the
low and large $\theta_n$  values. The plot of the local bending angles of
the 10 best conformations, seen in the insert in
Fig.\ref{fig:summary}-c, shows that they may have more than one large angle.
The findings are consistent with radiation damage to the DNA, which creates
single-strand breaks that allow local sites to be far more flexible.  
However, as for
the search at $23^{\circ}$C, a high concentration of large angles is
still found around the middle of the model with an angle distribution centered
at $\theta_n \sim 125^{\circ}$.  The number of sites within this distribution
is greater at $70^{\circ}$C which may represent more kinks due to increased
thermal fluctuations or large bendings due to single-strand breaks, which is
consistent with X-ray radiation damage preferentially causing single-strand
breaks at sites with local distortion \cite{MILLER-JH}, in this case due to
kinks.  Both observations indicate that kinks are present in Widom-601. 

\section{Complementary experiment with another DNA sequence}
\label{sec:complementary}

As a further test to validate our results, we have performed a complementary
experiment with a DNA sequence previously studied in cyclization experiments
and found to be well described by the standard WLC model of DNA\cite{QUANDU}.
This sequence, of 204 base pairs, is shown in Fig.~\ref{fig:seqsans1}. It is
derived from a segment of 200 base pairs of $\lambda$-phage DNA, starting at
site 29853, modified at its ends to make the PCR step easier in the sample
preparation. This sample had been chosen for the cyclization studies because it
does not have intrinsic curvature. However, in order to check the ability of
our experiments and analysis to detect specific features of the DNA molecules
under study, we modified the sequence locally, changing the some $G-C$ pairs
into $A-T$ in order to create a domain of 32 consecutive $A-T$ pairs from
position $92$ to position $125$. Due to the larger fluctuations of the $A-T$
pairs \cite{CUESTA} we expect this region to have a higher
flexibility. Moreover it includes some elements, such as AATT, which have a
small intrinsic curvature.

A solution with a concentration of $1.27\;$mg/mL was studied by
SANS, using the D22 experiment at Institut Laue Langevin as described
for the Widom-601 sequence. The scattering data were treated by the {\tt GNOM}
program to compute $P\left(r\right)$ and then we performed a computational
search using the same program as for the Widom-601 sequence and the same model
parameters, except for the number of nodes.
\begin{figure}[h]
  \begin{tabular}{c}
\includegraphics[width=7.3cm,clip]{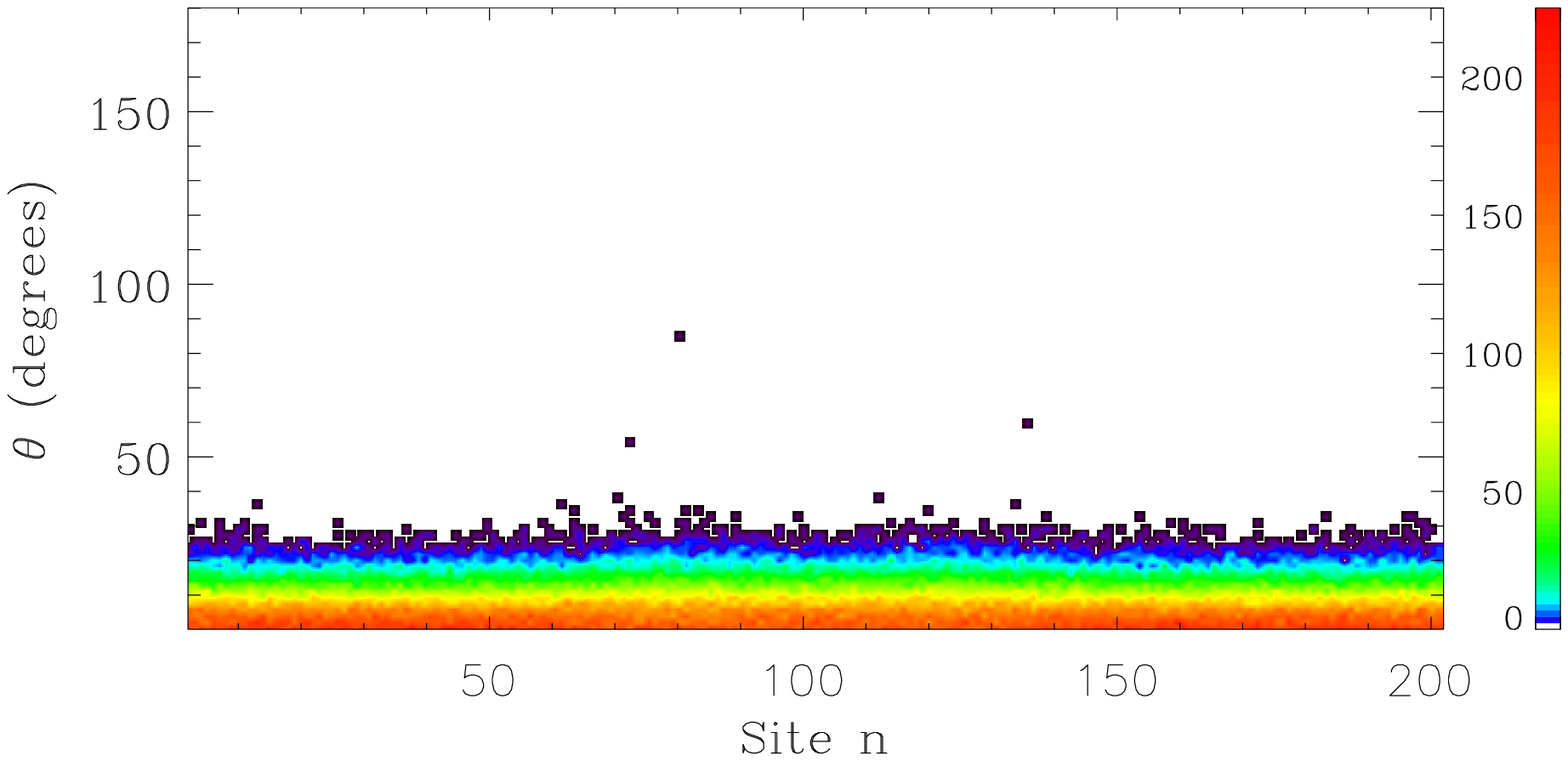}
\hspace{-0.3cm}
\includegraphics[height=3.5cm,clip]{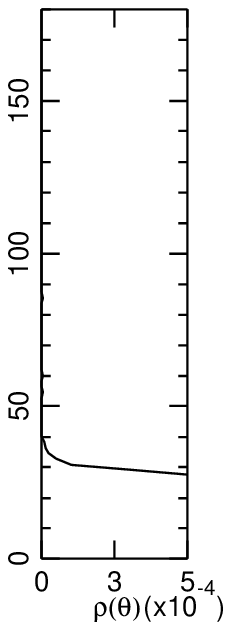}
\end{tabular}
\caption{Equivalent of Figs~\ref{fig:summary}-b and \ref{fig:summary}-d,
for the complementary experiment with the sequence
shown in Fig.~\ref{fig:seqsans1}. Histogram of the bending angle
against $n$ for the 10$^3$ conformations that provide the best matching with
the room temperature SANS data.
For each site of the polymer model, the left part shows the number of
 $\theta_n$ values that correspond to the value marked on the left scale (the
 total of these numbers for a given $n$ is equal to 1000, the number of
 conformations) with a color scale shown on the right. The right part of the
 figure shows the fraction $\rho$ of $\theta$ angles, integrated over the whole
 model, which belongs to a given range of theta. The scale is truncated to
$\rho_{\mathrm{max}} = 5 \times 10^{-4}$ to better show the part of the curve which
corresponds to large $\theta$ angles.}
  \label{fig:mcsans1}
\end{figure}

Figure \ref{fig:mcsans1} shows the histogram of the bending angles of the 
the 10$^3$ conformations that provide the best matching with
the room temperature SANS data, and its right part shows the fraction $\rho$
of $\theta$ angles, integrated over the whole 
 model, which belongs to a given range of theta. This figure is analogous to
Figs.~\ref{fig:summary}-b and \ref{fig:summary}-d 
for the Widom-601
sequence, and we used the same scales for the plots to allow a quantitative
comparison between the results for the two sequences. The difference is
striking. For the sequence taken from $\lambda$-phage DNA we do not find the
many kinked conformations detected for the nucleosome positioning sequence.

\begin{figure}[h]
\vspace{-0.8cm}
  \centering
  \includegraphics[width=6.8cm]{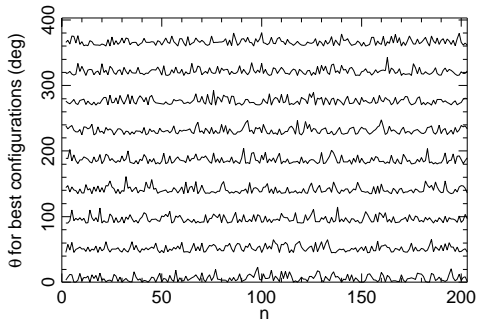}
\vspace{-0.8cm}
  \caption{Values of $\theta_n$ for the 10 conformations which provide
the best match to the experimental $P\left(r\right)$ deduced from
SANS measurements at room temperature on the sequence
shown in Fig.~\ref{fig:seqsans1} . Successive plots are moved up by
$50^{\circ}$ to limit the overlap between the curves. }
  \label{fig:valthetasans1}
\end{figure}
Figure \ref{fig:valthetasans1} for the sequence taken from $\lambda$-phage DNA
is equivalent to the inset on Fig.~\ref{fig:summary}-a
for the Widom-601 sequence. It shows
the values of the $\theta_n$ for the 10 conformations which provide
the best match to the experimental $P\left(r\right)$ deduced from
SANS measurements. Contrary to the case of the Widom-601 nucleosome
positioning sequence, the bending angles do not show any sharp spike,
associated to the presence of a kink.

\begin{figure}
  \includegraphics[width=7cm]{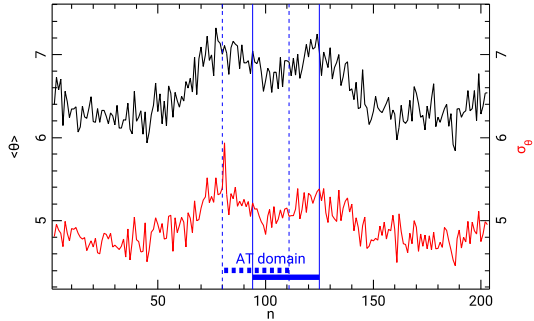}
  \caption{Average values the the local bending angles $\langle \theta_n
    \rangle$ and standard deviations of the bending angles for the 1000 
conformations which provide
the best match to the experimental $P\left(r\right)$ deduced from
SANS measurements at room temperature on the sequence
shown in Fig.~\ref{fig:seqsans1}. The blue lines show the position of the $AT$
domain along the sequence, starting either from the left end of the sequence
(full lines) or from the right end (dashed lines).  }
  \label{fig:thetasigma}
\end{figure}
The absence of kinks for this DNA sequence does not mean that the results of
the SANS experiments are featureless. To detect the specificities of this
sequence one has to look at the average bending angles $\langle \theta_n
\rangle$, and their standard
deviations $\sigma_{\theta}$. 
They are plotted in Fig.~\ref{fig:thetasigma}. The averages and
standard deviations have been calculated over the 1000 conformations which
provide the best matching with the experiments. Both the average bending and
its fluctuations show maxima in positions corresponding to the domain of pure
$AT$ pairs. As the measurements do not distinguish the two ends of the
molecules, the contribution of this domain on the data shows up simultaneously
on the sites $94$ to $125$ and on $111$ to $80$, which correspond to its
position measured from the other end of the sequence. Moreover, as we have
showed using another experimental approach\cite{CUESTA}, the fluctuations of a
large $AT$ domain influence the local conformation of DNA in its
vicinity. Therefore we expect to detect the influence of the $AT$ domain also
in its vicinity. This is exactly what the analysis of the SANS data detects,
as shown in Fig~\ref{fig:thetasigma}. The fluctuations of the $AT$ domain,
which lead to disturbance in base stacking locally reduce the bending rigidity
of double-stranded DNA. This leads to an increased standard deviation of
$\theta_n$ and an increase in its average value due to entropic effects,
which, in this case, are also reinforced by the small intrinsic curvature of
some $AT$ elements. Our measurements also detect the small increase in
$\langle \theta_n \rangle$ and $\sigma_{\theta}$ due to the free
ends. Therefore Fig~\ref{fig:thetasigma} shows that our measurements and their
analysis are able to detect fairly small effects in the conformation of DNA in
solution. This reinforces our statement about the existence of kinks in the
Widom-601 positioning sequence, which
are large distortions which should be easier to detect.

\section{Discussion}

Our experimental results on the Widom-601 sequence and their analysis clearly
point to a positive answer to the question raised by Crick and Klug in 1975
\cite{CRICK-KLUG}. Some DNA sequences can exhibit kinks, even in the absence
of strong external constraints.

A first hint was provided by the analysis of the data with standard software
packages developed for SAS data analysis. The fit by SASVIEW
\cite{SASVIEW}, as well as the  Kratky-Porod model \cite{KRATKY}, lead to
persistence lengths $l_p$ of the order of $100\;$\AA, which, at a first glance
appear unrealistically low. However this would be the case if $l_p$ was only
determined by dynamical fluctuations. But intrinsic curvature can also
contribute to reduce the effective persistence length \cite{SCHELLMAN}. This
effect was recently studied in details for various DNA sequences with an
elaborate coarse-grain DNA model \cite{MITCHELL} and it was found that it can
bring a significant contribution. Nevertheless, for the Widom-601 sequence,
our experiments indicated that the effect had to be quite dramatic to reduce
$l_p$ so much. To proceed further and determine the main features of the
molecular shapes from the data, we used an extended Kratky-Porod model in two
stages. First, an unbiased sampling of the conformational space using generic
model parameters, without sequence dependence or
intrinsic curvature, showed that the conformations providing the best match
with the data exhibited a sharp, highly localized bend, in their central
region. Moreover the statistics of the bending angles found a hump at large
angle, separated by a gap from the large peak around $\theta = 0$. This rules
out a highly flexible point caused for instance by a nick because it would
lead to a single-peaked distribution. The second stage, assuming
a non-zero equilibrium value of the bending angle at a particular site
showed that kinked conformations can indeed provide a good fit of the data.
The accuracy of the SAS experiments and of our analysis cannot formally rule
out a bending distribution extending over a few sites instead of a kink. This
is however very unlikely because, owing to the very large overall bending
required to fit the data, it would need several consecutive bends of
$20^{\circ}$ 
to $30^{\circ}$ which could hardly be achieved without fully breaking the DNA
structure with a high energetic cost. This is precisely because they
considered such a configuration as unlikely that Crick and Klug
\cite{CRICK-KLUG} looked for an alternative.
Instead, as shown by the model that they built, a kink in DNA can exist while
leaving all 
base-pair intact and all bond distances and angles stereochemically
acceptable.

Our analysis relies on the choice of a particular DNA model. We opted for a
model which is as simple as possible but contains the essential features
required to describe the DNA backbone. The Kratky-Porod model, often used, does
not include the dihedral energy. We added this term in the Hamiltonian because
it is important to control the overall shape of the polymer, which is probed
in SAS experiments. Dihedral energy prevents the free rotation about the
bonds, which could lead to large shape changes without affecting the bond-angle
energy which enters in the Kratky-Porod model. For the first stage of our
analysis, the search for conformations that best fit the data, it is important
to avoid any bias in the exploration of the conformational space, and
therefore the model has to be generic. The price to pay is that we have to
generate a huge number of conformations ($\approx 10^8$) to make sure that
they include those of the molecules in solution. This price is however
bearable because the model is sufficiently simple to allow fast calculations.
In the second stage of the analysis, we try instead to design a specific model
for the molecules in solution. Even within the extended Kratky-Porod model,
the number of parameters that could be adjusted is very large. We minimized
the number of free parameters by focussing our attention on a few equlibrium
values of the bending angles, as suggested by the conformational
search. Increasing the number of adjusted parameters could improve the
agreement with experiments, at the risk of ``over-fitting'' with parameters
that would not be statistically significant. Nevertheless it might be
interesting to refine our analysis with improved models for DNA, such as the
one used in \onlinecite{MITCHELL}. To cover all possible conformations, the
model would have to be parametrized to allow the description of kinks and not
only moderate local bending. The validity of our analysis has neverteless been
tested in a complementary experiment with another sequence.
Using the same
method of analysis we did not find kinks in this sequence
but demonstrated that our approach is able
to detect a fairly small effect in the conformation of DNA, validating the
method. The radiation dammage, caused by a long-exposure to X-Rays, provided
another, unexpected, complementary experiment. The analysis of the data is
able to detect the single stand breaks, which create additional flexible
points along the sequence, and also modify the probability distribution of the
bending angles, by removing the gap between the small angles and the very
large angles characteristic of a strong permanent bend.

Kinks in DNA are not new. They were suggested by the analysis of some
cyclization experiments \cite{CLOUTIER2005}, or detected in molecular dynamics
simulations of mini-circles \cite{LANKAS} and in the structure of the
nucleosome core particle \cite{VASUDEVAN}. {\em However all these examples
concerned highly constrained DNA. Our results show that kinks can also exist
for DNA samples in solution without any particular constraint}. Therefore this
peculiar DNA structure, proposed from a model building approach by Crick and
Klug, could be more common than generally assumed. However this is not a
generic property of DNA. Kinks depend on the sequence and appear to be present
in the nucleosome-positioning Widom-601 but not in a modified $\lambda$-page
sequence. Our findings may also help to resolve the recent debate concerning
the flexibility of short-chain DNA, which has focused on differences in
experimental protocols \cite{QUANDU} but which should also consider the
intrinsic properties of the DNA sequences that were investigated.

%

\begin{acknowledgments}

\bigskip
We thank the 15ID-D USAXS beamline at the Advanced Photon Source, USA,
who kindly provided the glassy carbon standard sample used for the absolute
normalization of the SAXS data.
T.S. and T.U. gratefully acknowledge the financial support by the Deutsche
Forschungsgemeinschaft (DFG) through the Cluster of Excellence “Engineering of
Advanced Materials” (EAM) and GRK 1896 “In-Situ Microscopy with Electrons,
X-rays and Scanning Probes”.
A.G. thanks the PhD program of the ILL for providing the financial support for
his thesis.
M.M.R., L.R.S. and S.C.L. acknowledge the overall 
support by the Spanish Ministry of
Economy, Industry and Competitiveness
(BES-2013-065453, EEBB-I-2015-09973, FIS2012-38827). S.C.L. and UC-154 are
grateful for the support of Junta de Castilla y Leon (Spain)
Nanofibersafe BU079U16. D.A. acknowledges funding
from the Agence Nationale de la Recherche through ANR-12-BSV5-0017-01
''Chrome'' and ANR-17-CE11-0019-03 ''Chrom3D'' grants.
N. T. acknowledges support by the project “Advanced Materials and
Devices” (MIS 5002409, Competitiveness, Entrepreneurship and Innovation,
NSRF 2014-2020) co-financed by Greece and the European Regional Development
Fund.

\end{acknowledgments}

\end{document}